\documentclass[aps,prl,superscriptaddress,twocolumn,balancelastpage,nofootinbib]{revtex4-2}

\usepackage[colorlinks,bookmarks=false,citecolor=blue,linkcolor=blue,urlcolor=blue]{hyperref}
\usepackage[all]{hypcap}   % let hyperlinks correctly point to figures rather than their captions;

\usepackage{amsmath,amssymb}
\usepackage{graphicx}

\usepackage{verbatim}
\usepackage{color}

\usepackage{placeins}    % for FloatBarrier
\usepackage{flafter}     % Bilder immer nach \figure Befehl
\usepackage{color}

\usepackage[normalem]{ulem}

\usepackage[capitalise]{cleveref}
\crefname{appendix}{}{}
\crefname{section}{Sec.}{}

\newcommand{\HIDDEN}[1]{}

\newcommand{\ue}{\text{e}}

\makeatletter
\let\Hy@backout\@gobble
\makeatother
\begin{document}

\title
{Semiclassical Limit of Resonance States in Chaotic Scattering}

\author{Roland Ketzmerick}
\affiliation{TU Dresden,
 Institute of Theoretical Physics and Center for Dynamics,
 01062 Dresden, Germany}

\author{Florian Lorenz}
\affiliation{TU Dresden,
 Institute of Theoretical Physics and Center for Dynamics,
 01062 Dresden, Germany}

\author{Jan Robert Schmidt}
\affiliation{TU Dresden,
 Institute of Theoretical Physics and Center for Dynamics,
 01062 Dresden, Germany}

\date{\today}

\begin{abstract}
Resonance states in quantum chaotic scattering systems have a multifractal
structure that depends on their decay rate.
We show how classical dynamics
describes this structure for all decay rates in the semiclassical limit.
This result for chaotic scattering systems corresponds to the well-established
quantum ergodicity for closed chaotic systems.
Specifically, we generalize Ulam's
matrix approximation of the Perron-Frobenius operator,
giving rise to conditionally
invariant measures of various decay rates.
There are many matrix approximations leading to the same decay rate
and we conjecture a criterion for selecting
the one
relevant for resonance states.
Numerically, we demonstrate that resonance states in the semiclassical limit
converge to the selected measure.
Example systems are a dielectric cavity, the three-disk scattering system,
and open quantum maps.

\end{abstract}

\maketitle

%%%%%%%%%%%%%%%%%%%%%%%%%%%%%%%%%%%%%%%%%%%%%%%%%%%%%%%%%%%%%%%%%%%%%%%%%%%%%
% Introduction
%%%%%%%%%%%%%%%%%%%%%%%%%%%%%%%%%%%%%%%%%%%%%%%%%%%%%%%%%%%%%%%%%%%%%%%%%%%%%

%%%%%%%%%%%%%%%%%%%%%%%%%%%%%%%%%%%%%%%%%%%%%%%%%%%%%%%%%%%%%%%%%%%%%%%%%%%%%
\emph{Introduction}---%
%%%%%%%%%%%%%%%%%%%%%%%%%%%%%%%%%%%%%%%%%%%%%%%%%%%%%%%%%%%%%%%%%%%%%%%%%%%%%
The structure of eigenstates in closed quantum systems,
which in the classical limit are ergodic,
is described by the semiclassical eigenfunction
hypothesis~\cite{Ber1977b, Vor1979, Ber1983}
and the quantum ergodicity
theorem~\cite{Shn1974, CdV1985, Zel1987, ZelZwo1996, BaeSchSti1998, NonVor1998}.
In the semiclassical limit almost all eigenstates
are uniform on the energy surface in phase space.

\begin{figure}[b!]
	\includegraphics[scale=1.0]{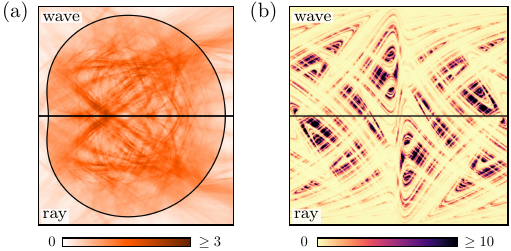}
	\caption{
		(a)
        Visual comparison between the average over 500 resonance states
        of a dielectric cavity of lima\c{c}on shape
        near $\mathrm{Re} \, kR_\text{cav} = 5000$
        with decay rate near $\gamma = 0.053$ (top)
        and the proposed conditionally invariant measure based on ray dynamics
        (bottom).
        (b)
        Comparison on boundary phase space.
   	}
	\label{FIG:qualitative_comparison}
\end{figure}

In contrast, in quantum scattering systems
with chaotic dynamics in the classical
limit~\cite{Smi1989, Gas1998, LaiTel2011, DyaZwo2019},
the structure of resonance states is much more
complex~\cite{CasMasShe1999b, KeaNovPraSie2006, AltPorTel2013},
see Fig.~\ref{FIG:qualitative_comparison}.
A recent factorization conjecture states that resonance states
are composed of a universal factor given by
a complex Gaussian random wave model
and a factor of classical origin giving a multifractal structure
depending on the decay rate~\cite{ClaKunBaeKet2021, KetClaFriBae2022, SchKet2023}.
Indeed, it was proven by Nonnenmacher and Rubin
that in the semiclassical limit
chaotic resonance states
are described by some
conditionally invariant measures~\cite{NonRub2007}.
Such measures are invariant under classical dynamics
up to a change of their norm due to escape~\cite{PiaYor1979, DemYou2006}.

Which are the conditionally invariant measures
corresponding to a quantum scattering system?
For resonance states close to one specific decay rate,
the so-called natural decay rate, the answer
is given by the natural measure~\cite{CasMasShe1999b}.
It is the eigenvector corresponding to the leading eigenvalue of the
Perron-Frobenius operator,
which describes the time evolution of densities
in phase space~\cite{CviArtMaiTan2020}.
The natural measure has been used extensively in dielectric cavities
to describe lasing
modes~\cite{LeeRimRyuKwoChoKim2004, ShiHar2007, LebLauZysSchBog2007, WieHen2008,
ShiHenWieSasHar2009, ShiHarFukHenSasNar2010, HarShi2015, CaoWie2015, KulWie2016,
BitKimZenWanCao2020, KetClaFriBae2022, KimBitJinZenWanCao2023}.
For such systems with partial escape
one can describe resonance states of a second decay rate,
the so-called inverse natural decay
rate, by the inverse natural measure~\cite{AltPorTel2015, GutOsi2015, ClaAltBaeKet2019}.
Poles of resonance states at the natural decay rate appear in the complex energy plane
close to the upper end of the spectrum,
while for the inverse natural decay they appear
close to the lower end~\cite{ClaAltBaeKet2019, KetClaFriBae2022}.

Much less is known for resonance states of any other decay rate~\cite{Nov2013}.
For systems with full escape, e.g.\ the three-disk scattering system, the support of resonance states is given by invariant sets of the classical dynamics~\cite{KeaNovPraSie2006, She2008, Non2011, Nov2013, WeiBarKuhPolSch2014, Dya2019, SchKet2023}.
Furthermore, for quantum maps with full escape
the measure in the opening
(and its preimages)
and how it depends on the decay rate was
derived in Ref.~\cite{KeaNovPraSie2006}.
The structure of resonance states was also related to short periodic
orbits~\cite{NovPedWisCarKea2009, ErmCarSar2009, PedWisCarNov2012,
CarBenBor2016, MonCarBor2024},
zeta functions~\cite{BarSchWei2022,SchWeiBar2023},
and finite-time Lyapunov exponents~\cite{ReySigPraSan2024}.
Another object of interest are Schur vectors determined from resonance
states~\cite{SchTwo2004, KopSch2010} which have been described by classical
densities~\cite{HalMalGra2023}.

There are approximate heuristic approaches for finding
conditionally invariant measures
describing resonance states,
which are fundamentally different for systems with full escape~\cite{ClaKoeBaeKet2018, SchKet2023}
and partial escape~\cite{ClaAltBaeKet2019, KetClaFriBae2022}.
Resonance states in the semiclassical limit come close, but do not converge, to these measures~\cite{ClaAltBaeKet2019, SchKet2023}.
So it remains an open question which are the
conditionally invariant measures corresponding to
the semiclassical limit of
chaotic resonance states of all decay rates.

In this Letter we construct the conditionally invariant measure
that describes the structure of chaotic resonance
states of a given decay rate in the semiclassical limit.
To this end we generalize Ulam's
matrix approximation of the Perron-Frobenius operator.
There are many matrix approximations
and we conjecture a criterion for selecting
the one
relevant for resonance states,
giving the desired conditionally invariant measure.
Numerically, we demonstrate
that resonance states in the semiclassical limit converge to the selected measure.
A visual comparison is shown for a dielectric cavity (partial escape)
in Fig.~\ref{FIG:qualitative_comparison}.
Further example systems are
the three-disk scattering system (full escape),
and quantum maps with partial and full escape.

%%%%%%%%%%%%%%%%%%%%%%%%%%%%%%%%%%%%%%%%%%%%%%%%%%%%%%%%%%%%%%%%%%%%%%%%%%%%%
\vspace*{0.1cm}
\emph{Ulam's method}---%
%%%%%%%%%%%%%%%%%%%%%%%%%%%%%%%%%%%%%%%%%%%%%%%%%%%%%%%%%%%%%%%%%%%%%%%%%%%%%
The time evolution of densities
in the phase space of a dynamical system
is governed by the Perron-Frobenius
operator~\cite{CviArtMaiTan2020}.
A matrix approximation of the operator goes back to
Ulam~\cite{Ula1960} and
has found many
applications~\cite{Fro1998, KeaMurYou1998, FroPad2009, FraShe2010, FraShe2013, ChaTanLoeSoe2013, YosYosShuLip2021}.
In the context of scattering systems
it was applied to maps with full escape~\cite{ErmShe2010}
and dielectric cavities~\cite{KulWie2016}.
There it generates the natural conditionally invariant measure
with the natural decay rate.
We first describe this approach, often called Ulam's method, which is
later generalized to arbitrary decay rates.

One partitions phase space into $n$
disjoint cells $\{ A_1, \dots, A_n\}$,
typically a grid of equally sized boxes.
For simplicity, we first consider a time-discrete map~$T$
on phase space.
It defines for
each cell $A_i$ subregions $A_{ji} \subset A_i$,
which are mapped
to cells $A_j$ under~$T$.
In other words, each subregion $A_{ji}$ is defined by
\begin{equation}
    \label{eq:subregion}
    A_{ji} = A_i \cap T^{-1}(A_j)
    \; .
\end{equation}
One defines a transition matrix (or stochastic matrix),
\begin{equation}
    \label{eq:transition_matrix_lebesgue}
    P_{ji}^{\mathcal{L}} = \frac{\mu_{\mathcal{L}}(A_{ji})}{\mu_{\mathcal{L}}(A_i)}
    \; ,
\end{equation}
where $\mu_{\mathcal{L}}$ is the uniform Lebesgue measure
on phase space.
We denote it as the \textit{Lebesgue transition matrix}~$P^{\mathcal{L}}$
to distinguish it from more general transition
matrices~$P$ to be introduced below.
Each element is the ratio of
the Lebesgue measure of the subregion $A_{ji}$
to the Lebesgue measure of the whole cell $A_i$.
From the above definitions follows
$\sum_j P^{\mathcal{L}}_{ji} = 1$,
as required for a transition matrix.
The matrix $P^{\mathcal{L}}$
is Ulam's matrix approximation of the Perron-Frobenius operator.
Numerically, it is determined by the fraction of trajectories
uniformly started within cell $A_i$ that goes to cell $A_j$.
We mention that for closed systems
the right leading eigenvector,
$\sum_i P^{\mathcal{L}}_{ji}\, \mu_i =  \mu_j$,
gives a coarse-grained invariant measure
$\mu_i = \mu(A_i)$
for all cells $A_i$ of the partition~\cite{Ula1960, CviArtMaiTan2020}.
Note that we use a notation for the order of indices
common in physics,
as, e.g., in Refs.~\cite{ErmShe2010, KulWie2016}.

In a scattering system with partial escape
one approximates the reflectivity
by a matrix $R$. Each element $R_{ji}$
is representative for the reflectivity of the transition $A_i \rightarrow A_j$.
In a system with full escape some of the $R_{ji}$ will be zero.
Combining a transition matrix $P$
with the reflectivity matrix $R$
by elementwise multiplication
leads to a matrix $P_{ji} R_{ji}$ with the eigensystem,
\begin{equation}
    \label{eq:eigen}
    \sum_i P_{ji} R_{ji} \, \mu_i
    =
    \ue^{-\gamma} \, \mu_j
    \; .
\end{equation}
Here the right leading eigenvector
defines a \emph{coarse-grained conditionally invariant measure} $\mu_i = \mu(A_i)$
for all cells $A_i$ of the partition.
The leading eigenvalue $\ue^{-\gamma}$ gives the
decay rate $\gamma$ of this measure.
Note that $\mu_i \ge 0$ for all $i$ is ensured by the
Perron-Frobenius theorem~\cite{BriStu2002} for the leading eigenvector.
Subleading solutions have positive and negative
entries. Thus they cannot be interpreted as measures and are of no relevance here.
The above approach was introduced in Ref.~\cite{ErmShe2010}
for maps with full escape
using the Lebesgue transition matrix
$P^{\mathcal{L}}$.

For time-continuous systems
one also has to consider the transition time~\cite{AltPorTel2013},
e.g.\ between reflections with a billiard boundary,
while in Eq.~\eqref{eq:eigen} the transition time
was implicitly set to 1.
The transition time is approximated by a matrix $t$, where
each element $t_{ji}$ is representative for the
transition \mbox{$A_i \rightarrow A_j$}.
In this case, the exponential factor in Eq.~\eqref{eq:eigen} has to be placed on the left-hand side,
yielding the eigensystem,
\begin{equation}
    \label{eq:eigen_true_time}
    \sum_i P_{ji} R_{ji} \ue^{\gamma t_{ji}} \, \mu_i = \mu_j
    \; ,
\end{equation}
with leading eigenvalue $1$.
Here the unknown decay rate $\gamma$ has to be adjusted,
such that the leading eigenvalue of the matrix
$P_{ji} R_{ji} \ue^{\gamma t_{ji}}$
(multiplied elementwise)
is indeed~$1$.
This can be done iteratively~\cite{SM}.
This approach (without adjusting $\gamma$) was introduced in Ref.~\cite{KulWie2016}
for dielectric cavities
using the Lebesgue transition matrix
$P^{\mathcal{L}}$.

In the limit of increasingly finer partitions of phase space
the number of cells increases, $n \rightarrow \infty$.
One desires that in this limit the coarse-grained conditionally invariant measure converges,
which is mathematically an open question.
In general, the chosen partition and functional space influence the
convergence, see the recent Ref.~\cite{YosYosShuLip2021} and references
therein.
Numerically,
using the Lebesgue transition matrix $P^{\mathcal{L}}$
in Eqs.~\eqref{eq:eigen} or \eqref{eq:eigen_true_time}
one finds convergence
to the natural measure $\mu_{\text{nat}}$ and the
natural decay rate $\gamma_{\text{nat}}$,
respectively~\cite{ErmShe2010,KulWie2016}.
Note that the natural measure $\mu_{\text{nat}}$
is typically not determined
using a matrix approximation.
Instead a long-term time evolution of any smooth density with the
Perron-Frobenius operator is done, which
is numerically implemented by iterating trajectories
and their intensities~\cite{CasMasShe1999b,LeeRimRyuKwoChoKim2004,AltPorTel2013}.

%%%%%%%%%%%%%%%%%%%%%%%%%%%%%%%%%%%%%%%%%%%%%%%%%%%%%%%%%%%%%%%%%%%%%%%%%%%%%
\vspace*{0.1cm}
\emph{Transition matrices for arbitrary decay rates}---%
%%%%%%%%%%%%%%%%%%%%%%%%%%%%%%%%%%%%%%%%%%%%%%%%%%%%%%%%%%%%%%%%%%%%%%%%%%%%%
The Lebesgue transition matrix $P^{\mathcal{L}}$,
Eq.~\eqref{eq:transition_matrix_lebesgue},
is based~on the restriction
that the measure of the subregions $A_{ji}$ of a cell $A_i$ is
given by the uniform Lebesgue measure.
This assumption
limits the possible coarse-grained measures to the natural measure
decaying with~$\gamma_{\text{nat}}$.

In order to find conditionally invariant measures of other decay rates,
$\gamma\ne\gamma_{\text{nat}}$,
we now lift this restriction.
We allow for
transition matrices $P \ne P^{\mathcal{L}}$
without any restriction on the properties of the measure
on the subregions. We define more general transition matrices,
\begin{equation}
    \label{eq:transition_matrix}
    P_{ji} = \frac{\mu(A_{ji})}{\mu(A_i)}
    \; ,
\end{equation}
with arbitrary measures $\mu$ replacing
the uniform Lebesgue measure $\mu_{\mathcal{L}}$
in Eq.~\eqref{eq:transition_matrix_lebesgue}.
Note that
small deviations from uniformity were previously used in
proofs on the convergence for $n \rightarrow \infty$~\cite{Fro1998, KeaMurYou1998}.

The above definition,
Eq.~\eqref{eq:transition_matrix},
is equivalent to using all transition matrices $P$
(i.e.\ matrices with non-negative elements and $\sum_j P_{ji} = 1$),
which are
compatible with the allowed transitions between the cells of the partition encoded in $P^{\mathcal{L}}$, i.e.\
\begin{equation}
    P_{ji} \ne 0
    \quad \text{only if} \quad
    P^{\mathcal{L}}_{ji} \ne 0
    \; .
\end{equation}
In other words, for a given partition we allow for all
possible matrix approximations $P$ of the
Perron-Frobenius operator.

For each such transition matrix $P$
one finds from the leading eigenvector of Eq.~\eqref{eq:eigen} or \eqref{eq:eigen_true_time}
a coarse-grained conditionally invariant measure $\mu$ and its decay rate $\gamma$.
Decay rates occur over a wide range that depends on the
maximal and minimal entries of the reflectivity matrix~$R$, the transition times~$t$,
and the allowed transitions.
In particular, a given decay rate $\gamma$ from this range
is found for infinitely many transition matrices $P$,
which give rise to infinitely many different coarse-grained conditionally invariant measures~$\mu$.

%%%%%%%%%%%%%%%%%%%%%%%%%%%%%%%%%%%%%%%%%%%%%%%%%%%%%%%%%%%%%%%%%%%%%%%%%%%%%
\vspace*{0.1cm}
\emph{Selection criterion for transition matrix}---%
%%%%%%%%%%%%%%%%%%%%%%%%%%%%%%%%%%%%%%%%%%%%%%%%%%%%%%%%%%%%%%%%%%%%%%%%%%%%%
Which of the above infinitely many coarse-grained conditionally invariant measures
for a given decay rate $\gamma$
is relevant for describing quantum mechanical
resonance states with this decay rate?
We conjecture that it is the measure emerging from the
transition matrix $P$ that is \textit{closest} to
$P^{\mathcal{L}}$,
in the sense of minimizing the Kullback-Leibler divergence from
$P$ to $P^{\mathcal{L}}$,
\begin{equation}
    \label{eq:divergence}
    d(P^{\mathcal{L}} || P) = \frac{1}{n} \sum_{i} \,
    \left( -
    \sum_{j}
    P^{\mathcal{L}}_{ji} \, \ln
    \frac{P_{ji}}{P^{\mathcal{L}}_{ji}}
    \right)
    \; ,
\end{equation}
evaluated for each cell $A_i$ and averaged over all cells.
Its lowest value zero occurs for the special case $P=P^{\mathcal{L}}$,
which is consistent with $\gamma=\gamma_{\text{nat}}$.
For any other $\gamma\ne\gamma_{\text{nat}}$ one finds $d(P^{\mathcal{L}}||P)>0$.
Note that the Kullback-Leibler divergence depends on the order of its arguments
and that the chosen ordering uses the Lebesgue transition matrix $P^{\mathcal{L}}$
as the reference.

\begin{figure}
	\includegraphics[scale=1.0]{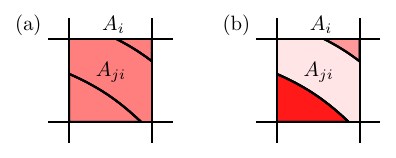}
        \vspace{-0.5cm}
	\caption{
        Cell $A_i$ of the partition
        and its subregions $A_{ji} \subset A_i$,
        Eq.~\eqref{eq:subregion}.
		(a)~Uniform density giving
        Lebesgue transition matrix~$P^{\mathcal{L}}$.
        (b)~Nonuniform density giving
        a general transition matrix~$P$.
   	}
	\label{FIG:subregions}
\end{figure}

We can derive this selection criterion for
\textit{locally randomized} scattering systems in the semiclassical limit
using a local random vector model in each cell~$A_i$~\cite{SM}.
It was introduced for the special case of the
randomized Baker map with escape,
perfectly describing resonance states of all decay rates~\cite{ClaKet2022}.
Note that
resonance states of the deterministic Baker map,
however, showed small deviations
from those of the randomized Baker map.
We attribute this to anomalies
due to the discontinuity of the Baker map.

For a better intuition it helps to
relate the elements $P_{ji}$ of a transition matrix
to a density distribution within a cell $A_i$,
see Fig.~\ref{FIG:subregions}.
For the Lebesgue transition matrix $P^{\mathcal{L}}$,
Eq.~\eqref{eq:transition_matrix_lebesgue},
the measure $\mu_{\mathcal{L}}(A_{ji})$
of each subregion $A_{ji}$ is obtained by
integrating a uniform density.
Instead, for a general transition matrix $P$,
Eq.~\eqref{eq:transition_matrix},
we allow for a nonuniform density within the cell.
Such $P$ give rise to conditionally invariant
measures with other decay rates.
The selection criterion chooses among the nonuniform densities
giving the desired decay rate,
the one closest to a uniform density.

Using the method of Lagrange multipliers we
select
the transition matrix $P$ closest to $P^{\mathcal{L}}$
under the constraints on $P$ and
with a measure~$\mu$ having the given decay rate~$\gamma$.
This leads
to a set of nonlinear equations~\cite{SM},
\begin{align}
    \label{eq:nonlinear_set_main_1}
    \phantom{\sum_j}
    P_{ji}
    &=
    \frac{P^{\mathcal{L}}_{ji}}{1 + \left( y_j R_{ji}  \ue^{\gamma t_{ji}}  - y_i \right) \mu_i}
    &\forall i,j
    \\
    \label{eq:nonlinear_set_main_2}
    \sum_j P_{ji}
    &=
    1
    & \forall i
    \\
    \label{eq:nonlinear_set_main_3}
    \sum_i P_{ji} R_{ji}  \ue^{\gamma t_{ji}} \, \mu_i
    &=
    \mu_j
    &\forall j
    \\
    \label{eq:nonlinear_set_main_4}
    \sum_i \mu_i
    &=
    1
    \; ,
\end{align}
with the unknown variables of interest,
$P_{ji}$ and $\mu_i$,
as well as the unknown Lagrange multipliers $y_i$.
Numerically, these nonlinear equations can be solved iteratively and we provide Python code~\cite{SM}.
The special case, $\gamma = \gamma_{\text{nat}}$,
leads to
$P = P^{\mathcal{L}}$, $\mu = \mu_{\text{nat}}$, and all $y_i = 0$.
In general for $\gamma \ne \gamma_{\text{nat}}$, one finds the elements $P_{ji}$ to be quite different
from $P^{\mathcal{L}}_{ji}$,
in some cases by large factors.

%%%%%%%%%%%%%%%%%%%%%%%%%%%%%%%%%%%%%%%%%%%%%%%%%%%%%%%%%%%%%%%%%%%%%%%%%%%%%
\vspace*{0.1cm}
\emph{Example systems}---%
%%%%%%%%%%%%%%%%%%%%%%%%%%%%%%%%%%%%%%%%%%%%%%%%%%%%%%%%%%%%%%%%%%%%%%%%%%%%%
We will use four example systems to demonstrate
for the selected measure
(i) the convergence in the limit $n \rightarrow \infty$
of finer partitions
and (ii) the agreement with resonance states in the semiclassical limit.
These examples cover the cases of partial and full escape
for a 2D billiard and a map in each case:
\begin{enumerate}
\item[(a)]
Dielectric cavity (partial escape) with lima\c{c}on shape,
deformation $\varepsilon = 0.6$, where it is practically fully chaotic,
radius $R_\text{cav}$,
refractive index $n_{\text{r}}=3.3$,
and reflection law for a TM polarized mode~\cite{KetClaFriBae2022}.
\item[(b)]
Three-disk scattering system (full escape)
with disks of radius $a$ and
center to center distance $2.1 a$~\cite{SchKet2023}.

\item[(c)]
Standard map at kicking strength $K=10$, where it is practically fully chaotic,
with partial reflectivity $0.2$
in the interval $q \in [0.3, 0.6]$~\cite{ClaAltBaeKet2019}.

\item[(d)]
Like (c) but with reflectivity 0 (full escape)~\cite{ClaAltBaeKet2019}.

\end{enumerate}
For details on these systems see~\cite{SM}.
We provide Python code to compute $P^{\mathcal{L}}$, $P$, and the proposed
measure $\mu$~\cite{SM}.
In all cases the elements of $P^{\mathcal{L}}$
are determined using $10^4$ trajectories per cell of the partition
(started on a uniform grid)
for up to $n = 3200^2 \approx 10^7$ cells.

\begin{figure}[h!]
   	\includegraphics{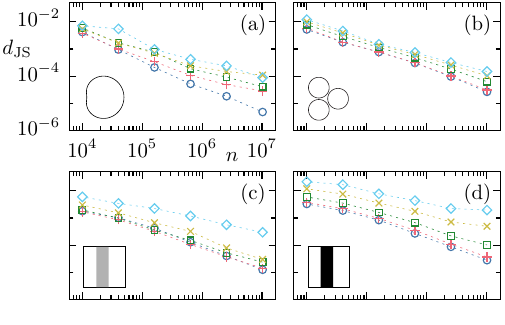}
	\caption{
        Convergence of coarse-grained conditionally invariant measures
        for increasingly fine partitions with $n$ cells.
        Shown is the
        decay of the Jensen-Shannon divergence $d_{\text{JS}}$ between
        measures from partitions $n$ and $n/4$.
        (a) Dielectric cavity for
        $\gamma \in \{ 0.011 \, (\gamma_{\text{nat}}), 0.030, 0.053, 0.090, 0.122 \, (\gamma_{\text{inv}}) \}$.
        (b) Three-disk scattering system for
        $\gamma \in \{0.436 \, (\gamma_{\text{nat}}), 0.6, 1.0, 1.4, 1.8 \}$.
        (c) Standard map with partial escape for
        $\gamma \in \{0.22 \, (\gamma_{\text{nat}}), 0.35, 0.55, 0.75, 0.88 \, (\gamma_{\text{inv}}) \}$.
        (d) Standard map with full escape for
        $\gamma \in \{0.25 \, (\gamma_{\text{nat}}), 0.35, 0.5, 0.75, 1.0 \}$.
        Symbols {\large $\circ$}, $+$, {\scriptsize $\square$}, $\times$, and {\large $\diamond$}
        are used for increasing $\gamma$.
    }
	\label{FIG:finer_partition}
\end{figure}

%%%%%%%%%%%%%%%%%%%%%%%%%%%%%%%%%%%%%%%%%%%%%%%%%%%%%%%%%%%%%%%%%%%%%%%%%%%%%
\emph{(i) Convergence in the limit $n \rightarrow \infty$}---%
%%%%%%%%%%%%%%%%%%%%%%%%%%%%%%%%%%%%%%%%%%%%%%%%%%%%%%%%%%%%%%%%%%%%%%%%%%%%%
We numerically demonstrate the convergence of the coarse-grained measure
in the limit $n \rightarrow \infty$ of a finer partition.
To this end we
compare the coarse-grained measures for $n$ and for $n/4$ cells
by integrating them on a $50 \times 50$ grid
and using the Jensen-Shannon divergence~\cite{Lin1991}
(its square root being a metric).
Figure~\ref{FIG:finer_partition}
shows that with increasing $n$ the Jensen-Shannon divergence
converges to zero.
Furthermore, the ratio
of two consecutive Jensen-Shannon divergences
is bounded from above by a value smaller than $1$.
If this continues to hold for increasing $n$,
the coarse-grained measures
are a contractive sequence,
therefore a Cauchy sequence and thus converge
in the limit $n \rightarrow \infty$.
This is demonstrated for various decay rates
$\gamma \in [\gamma_{\text{nat}}, \gamma_{\text{inv}}]$ (partial escape)
and
$\gamma \ge \gamma_{\text{nat}}$ (full escape)
for the billiard systems and maps.

\begin{figure}[h!]
	\includegraphics{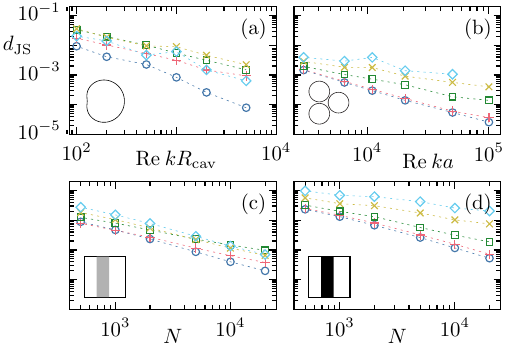}
	\caption{
        Convergence of averaged quantum resonance states
        to proposed conditionally invariant measures with $n = 3200^2 \approx 10^7$
        for systems and decay rates of Fig.~\ref{FIG:finer_partition}.
        Shown is the
        decay of the Jensen-Shannon divergence $d_{\text{JS}}$
        in the semiclassical limit, i.e.\
        (a, b) increasing
        wave number $\mathrm{Re} \, k$
        or (c, d)
        matrix size $N$.
        Details on the used resonance states are given in~\cite{SM}.
   	}
	\label{FIG:semiclassical_limit}
\end{figure}

%%%%%%%%%%%%%%%%%%%%%%%%%%%%%%%%%%%%%%%%%%%%%%%%%%%%%%%%%%%%%%%%%%%%%%%%%%%%%
\emph{(ii) Comparison with resonance states in the semiclassical limit}---%
%%%%%%%%%%%%%%%%%%%%%%%%%%%%%%%%%%%%%%%%%%%%%%%%%%%%%%%%%%%%%%%%%%%%%%%%%%%%%
We compare Husimi representations of
quantum resonance states near some decay rate $\gamma$
to the corresponding proposed measure from the finest partition.
In order to be most sensitive, we use the average over
500 resonance states of similar decay rate.
The visual comparison is demonstrated in
Fig.~\ref{FIG:qualitative_comparison}
for the dielectric cavity at one decay rate.
Note that resonance states show multifractality on scales larger than a
Planck cell only and we have to smooth the measure on this scale for the visual comparison
in Fig.~\ref{FIG:qualitative_comparison}(b).
For further examples see Supplemental Material~\cite{SM}.

Quantitatively,
Fig.~\ref{FIG:semiclassical_limit}
shows that going further towards the semiclassical limit
the Jensen-Shannon divergence (evaluated on a $50 \times 50$ grid)
converges to zero.
This is demonstrated for various decay rates
$\gamma \in [\gamma_{\text{nat}}, \gamma_{\text{inv}}]$ (partial escape)
and
$\gamma \ge \gamma_{\text{nat}}$ (full escape)
for the billiard systems and maps.
This convergence to zero
is in contrast to previous approximate approaches
for conditionally invariant measures~\cite{ClaKoeBaeKet2018, ClaAltBaeKet2019},
which describe resonance states quite well, but not
perfectly~\cite{ClaAltBaeKet2019, SchKet2023}.
We attribute different absolute values and convergence speeds
to properties of the multifractal resonance states.
These numerical results for various types of scattering systems
give strong support for the conjectured selection criterion
for the transition matrix and the corresponding measure.

%%%%%%%%%%%%%%%%%%%%%%%%%%%%%%%%%%%%%%%%%%%%%%%%%%%%%%%%%%%%%%%%%%%%%%%%%%%%%
\vspace*{0.1cm}
\emph{Remarks}---%
%%%%%%%%%%%%%%%%%%%%%%%%%%%%%%%%%%%%%%%%%%%%%%%%%%%%%%%%%%%%%%%%%%%%%%%%%%%%%
Let us stress that for individual resonance states we find just as well
a convergence to the proposed measure~\cite{SM}, as expected by the factorization conjecture~\cite{ClaKunBaeKet2021, KetClaFriBae2022, SchKet2023}.
However, one observes a much larger Jensen-Shannon divergence due to their fluctuations.
This makes individual resonance states a much less sensitive test
for the quality of a conditionally invariant measure~\cite{ClaAltBaeKet2019}.

The analysis is presented for right resonance states.
It can be straightforwardly extended to left resonance states
of billiards or maps with escape.
This should allow for predicting the structure of the left-right Husimi representation~\cite{ErmCarSar2009}.

Instead of conjecturing closeness of the transition matrix $P$
to $P^{\mathcal{L}}$, as proposed here,
one could conjecture closeness of $P$ to a transition matrix $P^{\text{nat}}$
given by Eq.~\eqref{eq:transition_matrix_lebesgue}
with $\mu_{\mathcal{L}}$ replaced by the
natural measure $\mu_{\text{nat}}$~\cite{Cla2024:p}.
By definition, this works for the natural decay rate.
Also for other decay rates we find in the limit
$n \rightarrow \infty$ convergence to the same measures as before.
An advantage of using $P^{\text{nat}}$ is that we observe
faster convergence in the limit $n \rightarrow \infty$.
A disadvantage is that one needs an approximation of the
natural measure $\mu_{\text{nat}}$ on a much finer scale than the
partition.

%%%%%%%%%%%%%%%%%%%%%%%%%%%%%%%%%%%%%%%%%%%%%%%%%%%%%%%%%%%%%%%%%%%%%%%%%%%%%
\vspace*{0.1cm}
\emph{Outlook}---%
%%%%%%%%%%%%%%%%%%%%%%%%%%%%%%%%%%%%%%%%%%%%%%%%%%%%%%%%%%%%%%%%%%%%%%%%%%%%%
It is desirable to find a
semiclassical derivation for our selection criterion
of the transition matrix leading to the proposed measures, as it is possible for locally randomized systems~\cite{SM}.
Numerically, it might be of interest to find an alternative method to compute the proposed measures based on iterating trajectories for long
times, as it is common for the natural measure.

Extremely long-lived resonance states with $\gamma < \gamma_{\text{nat}}$ exist
at finite wavelengths, but not in the semiclassical limit~\cite{Nov2013, GutOsi2015, SchKet2023}.
Correspondingly, we find coarse-grained conditionally invariant measures
for $\gamma < \gamma_{\text{nat}}$, however,
for finite number $n$ of cells only
and not in the limit
$n \rightarrow \infty$.
This regime
of long-lived resonance states will be studied in the future.

Furthermore, it would be interesting to extend the set of examples to
systems with scattering in smooth potential, e.g.,
models of the chaotic ionization of atoms and of resonances in chemical
reactions~\cite{RamPraBorFar2009, BucDelZak2002}.
Also the relation to the recently found properties of
Schur eigenstates~\cite{HalMalGra2023} needs to be investigated,
as well as
the relation to quantum and classical channels in bipartite many-body systems~\cite{VijLak2025}.

%%%%%%%%%%%%%%%%%%%%%%%%%%%%%%%%%%%%%%%%%%%%%%%%%%%%%%%%%%%%%%%%%%%%%%%%%%%%%
\acknowledgments

\emph{Acknowledgments}---%
We are grateful for discussions with
E.~Altmann,
A.~B\"acker,
S.~Barkhofen,
K.~Clau\ss,
T.~Harayama,
M.~Richter,
P.~Schlagheck,
A.~Shudo,
G.~Tanner,
T.~T{\'e}l,
and
T.~Weich.
Funded by the Deutsche Forschungsgemeinschaft (DFG, German Research
Foundation)---262765445.
The authors gratefully acknowledge the computing time on
the high-performance computer at the NHR Center of TU Dresden.
This center is jointly supported by the Federal Ministry of Education and
Research and the state governments participating in the NHR.

%%%%%%%%%%%%%%%%%%%%%%%%%%%%%%%%%%%%%%%%%%%%%%%%%%%%%%%%%%%%%%%%%%%%%%%%%%%%%

\end{document}